\documentclass[aps,amsfonts,prl,twocolumn,showpacs]{revtex4-1}

\usepackage{pdfpages} 
\makeatletter
\AtBeginDocument{\let\LS@rot\@undefined}
\makeatother

\usepackage{subfigure}
\usepackage{textcomp}
\usepackage{graphicx}
\usepackage{bm}
\usepackage{amsmath,amssymb,amsthm}
\usepackage{amsfonts}
\usepackage{dsfont}
\usepackage[colorlinks=true,linkcolor=blue,urlcolor=blue,citecolor=blue]{hyperref}
\usepackage{multirow}
\usepackage{url}

\def\parfig#1#2{
\parbox{#1\textwidth}
{\includegraphics[width=#1\textwidth]{#2}}
}

\begin{document}
\title{Emergence of Navier-Stokes hydrodynamics in chaotic quantum circuits}

\author{Hansveer Singh$^{1}$, Ewan McCulloch$^{2}$, Sarang Gopalakrishnan$^{2}$ and Romain Vasseur$^{1}$}
\affiliation{$^1$Department of Physics, University of Massachusetts, Amherst, Massachusetts 01003, USA \\
$^2$Department of Electrical and Computer Engineering, Princeton University, Princeton NJ 08544, USA\\
}

\begin{abstract}

We construct an ensemble of two-dimensional nonintegrable quantum circuits that are chaotic but have a conserved particle current, and thus a finite Drude weight. The long-wavelength hydrodynamics of such systems is given by the incompressible Navier-Stokes equations. By analyzing circuit-to-circuit fluctuations in the ensemble we argue that these are negligible, so the circuit-averaged value of transport coefficients like the viscosity is also (in the long-time limit) the value in a typical circuit. The circuit-averaged transport coefficients can be mapped onto a classical irreversible Markov process. Therefore, remarkably, our construction allows us to efficiently compute the viscosity of a family of strongly interacting chaotic two-dimensional quantum systems.

\end{abstract}

\maketitle

\emph{Introduction}.---
The long-wavelength, late-time dynamics of generic many-body systems is governed by hydrodynamics. 
One can argue based on general principles based on chaos or scrambling that hydrodynamics must eventually emerge. 
However, microscopic derivations of hydrodynamic transport coefficients, or even the timescales on which hydrodynamic behavior sets in, starting from unitary dynamics or even classical dynamics of an isolated many-body system, remain intractable in general~\cite{bonetto2000fouriers}.
%
It was realized only recently that the emergence of hydrodynamics can be derived microscopically in ensembles of random circuits~\cite{Khemani:2018aa,Rakovszky:2018aa,Friedman:2019aa,McCulloch:2023aa}. Random-circuit methods are inherently defined on the lattice, with the circuit randomness strongly breaking translation invariance. Therefore, momentum relaxes rapidly, and the resulting hydrodynamics is therefore generally diffusive, or subdiffusive if kinetic constraints or quenched randomness are present~\cite{Singh:2021aa,Feldmeier:2022aa,Feldmeier:2020aa,Iaconis:2021aa,Morningstar:2020aa,Moudgalya:2021aa,10.21468/SciPostPhys.14.6.140,Gromov:2020aa,Iaconis:2019aa,Guo:2022aa,Hart:2022aa}.%

The richer phenomenology of Navier-Stokes hydrodynamics, which has stimulated a great deal of experimental and theoretical work in the context of low-temperature transport in graphene and other ultraclean metals~\cite{Lucas_2018,fritz2023hydrodynamic}, might appear to require momentum conservation and thus not be tractable using random-circuit techniques (or even, more generally, using lattice models with finite local Hilbert-space dimension).

In this work, we construct a family of two-dimensional random unitary circuits, which we dub Frisch–Hasslacher–Pomeau (FHP) random unitary circuits, inspired by lattice gas automata which exhibits incompressible Navier-Stokes hydrodynamics ~\cite{Frisch:1986aa,lgahydro1990}. Although the presence of the lattice breaks translation symmetry and momentum conservation, we show that the resulting hydrodynamics still exhibits an emergent momentum conservation law. The behavior we find bears close similarities to emergent Navier-Stokes hydrodynamics in systems with polygonal Fermi-surfaces~\cite{Cook:2019aa,10.21468/SciPostPhys.14.5.137,Qi:2023aa}. The dynamics averaged over the ensemble of circuits is described by a classical Markov chain, and the averaged correlation functions are easy to compute classically. Furthermore, there have been numerous works on quantum lattice Boltzmann methods which effectively simulate classical Markov processes capturing fluid behavior using quantum circuits~\cite{palpacelli2008quantum}. However, our interest is in the dynamics of \emph{individual} circuits. One of our main technical contributions is to compute the circuit-to-circuit fluctuations of transport coefficients and show that these are subleading (and quantitatively negligible) at late times. Relying on this ``self-averaging'' result, 
we use the ensemble averaged dynamics to study transport coefficients in a typical circuit. We indeed find that model exhibits damped sound modes as well as a diffusive mode with a finite d.c.~shear viscosity (up to logarithmic corrections~\cite{Forster:1977aa}).
Our construction has two particularly notable features. First, it provides a class of interacting two-dimensional lattice systems that still possess a nonzero Drude weight. Second, it allows us to extract transport coefficients---including the viscosity of a strongly interacting two-dimensional quantum system---through efficient classical simulations.

\emph{FHP Circuit Rules}.---
\begin{figure}[!t]
    \includegraphics[width=\columnwidth]{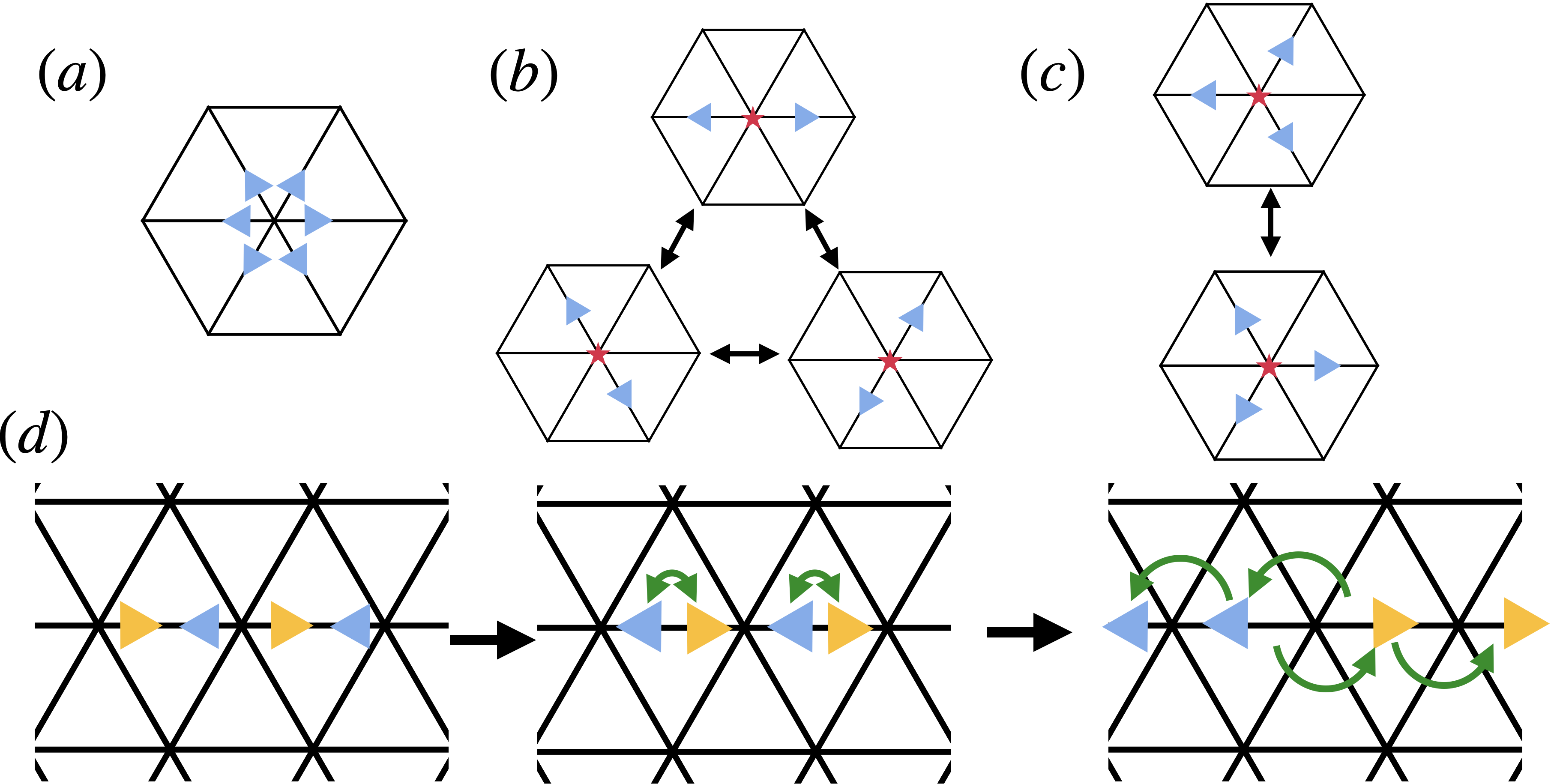}
    \caption{{\bf Hilbert Space and Circuit Rules.}(a) Degrees of freedom, represented as outwardly pointing arrows, reside on links incident to a site denoted by the red star. (b-c) Two and three body configurations, respectively, that the collision unitary has non-zero matrix elements. (c) Propagation step for the automaton achieved by two layers of SWAP gates. Particles moving to the left (right) are colored blue (yellow) here for clarity.  The first layer interchanges degrees of freedom associated with different nearest neighbor sites but reside on the same link. The second layer interchanges degrees of freedom associated with the same site but on different links along the same reflection axis of the hexagon.}
    \label{fig:fig1}
\end{figure}
The system resides on a two dimensional triangular lattice where each site, $\vec{x}$, hosts a Hilbert space, $\mathcal{H}_{\vec{x}}$, comprised of degrees of freedom living on links incident to site $\vec{x}$. More precisely,  $\mathcal{H}_{\vec{x}}$= $\otimes_{\ell} \mathcal{H}_{\vec{x}}^{\ell}$ where $\ell=1,...,6$ labels the six links incident to a given site $\vec{x}$ and 
\begin{equation}
\mathcal{H}^{l}_{\vec{x}} = \text{span}\{ |\parfig{.01}{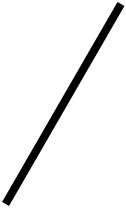}\rangle, |\parfig{.01}{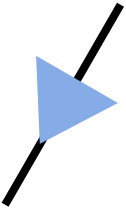}\rangle \}\otimes {\mathbb C}^{d}.
\end{equation}
This means each incident link to site $\vec{x}$ hosts a particle (represented as an arrow moving away from site $\vec{x}$ as shown in Fig.~\ref{fig:fig1}) and a $d$ dimensional ancillary degree of freedom to be used later for computing circuit to circuit fluctuations.\par 
The unitary operator generating the evolution $\mathcal{U}$ is split up into two parts: a collision step, whose unitary is denoted by $\mathcal{U}_C$, followed by a propagation step whose unitary is denoted by $\mathcal{U}_{P}$, i.e. $\mathcal{U} = \mathcal{U}_P\mathcal{U}_C$. The collision unitary operator $\mathcal{U}_C$ is comprised of unitary gates that act on each site at position $\vec{r}$, i.e. $\mathcal{U}_C = \prod_{\vec{r}} U_{C,\vec{r}}$. Each gate $U_{C,\vec{r}}$ has the following form
\begin{equation}
U_{C,{\vec{r}}} = P_{2} V_{2,\vec{r}} P_{2} + P_{3} V_{3,\vec{r}} P_{3} + \sum_{k}e^{i\phi_{k,\vec{r}}} P_{k},
\end{equation}
where the $P_a$ project onto the separately conserved sectors. The sectors with no collisions, collectively $\mathcal{H}_\perp$, are enumerated by an index $k$, i.e., $\mathcal{H}_\perp = \oplus_k \mathcal{H}_{\perp_k}$. Since these sectors are one dimensional, the Haar random unitary simply gives each of them a random phase $e^{i\phi_k}$. For the sectors with two and three body collisions, we have the projectors $P_a$ given by
\begin{align}
P_{2} &= \bigg|\parfig{.04}{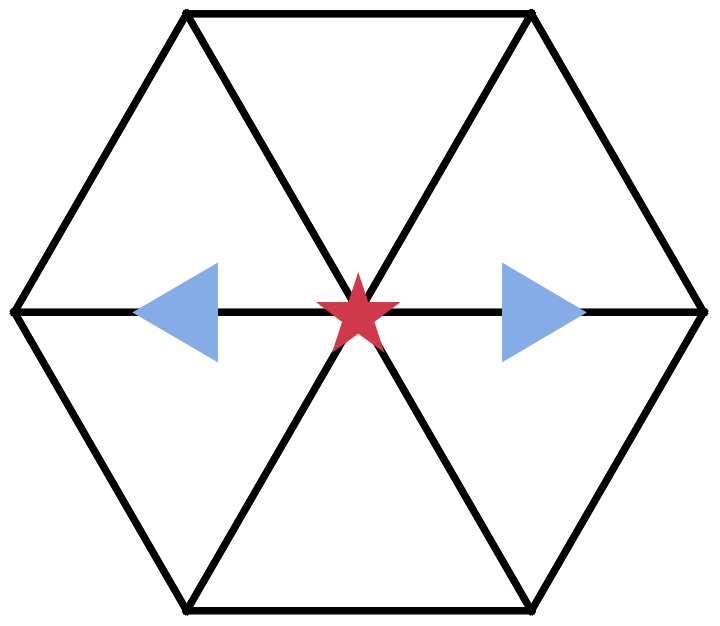}\bigg\rangle\bigg\langle\parfig{.04}{2body_1v3}\bigg|+\bigg|\parfig{.04}{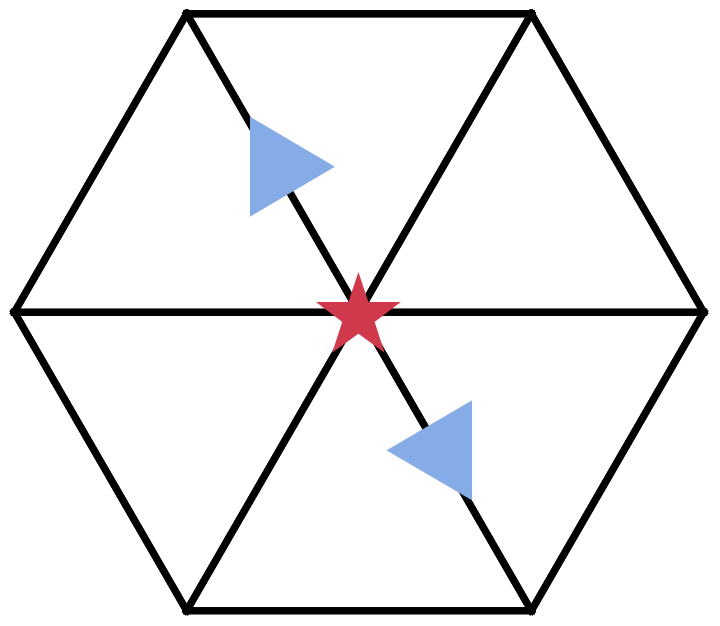}\bigg\rangle\bigg\langle\parfig{.04}{2body_2v3}\bigg|+\bigg|\parfig{.04}{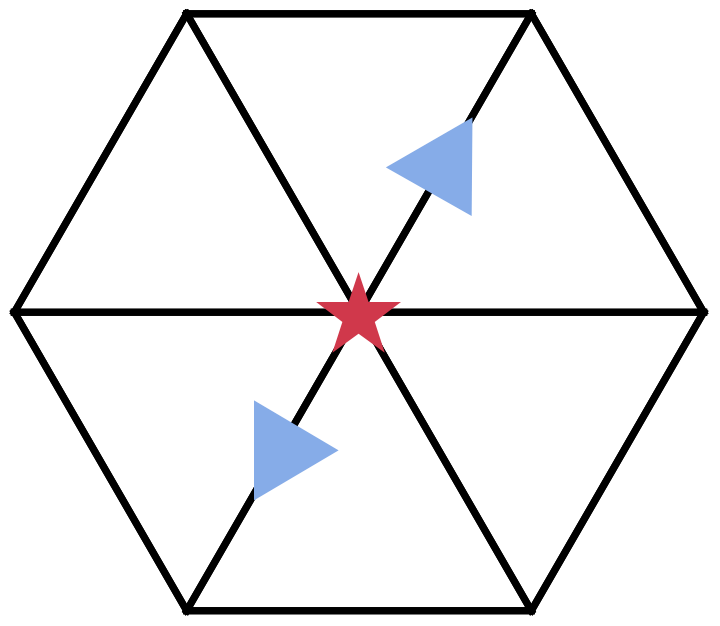}\bigg\rangle\bigg\langle\parfig{.04}{2body_3v3}\bigg|\\
P_{3} &= \bigg|\parfig{.04}{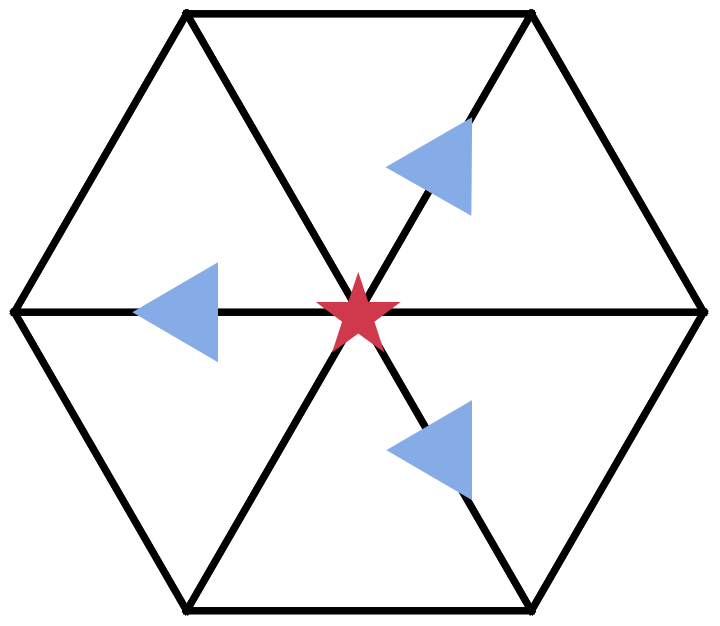}\bigg\rangle\bigg\langle\parfig{.04}{3body_1v3}\bigg|+\bigg|\parfig{.04}{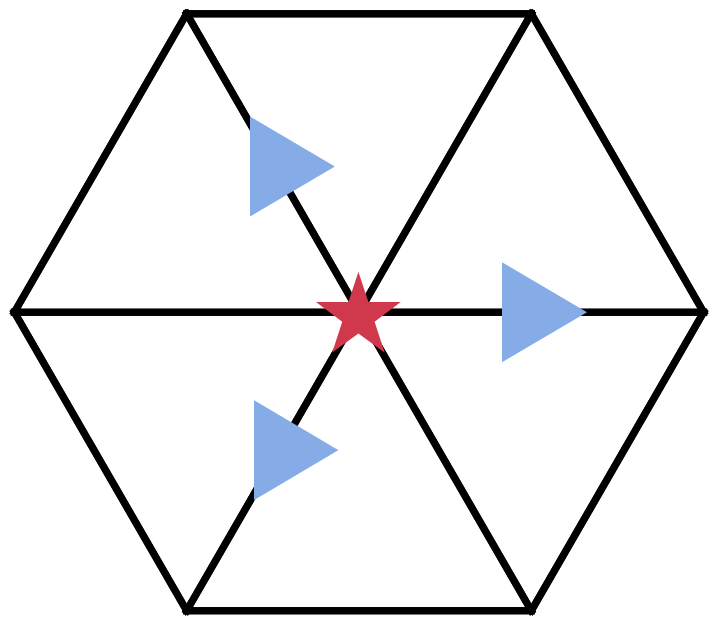}\bigg\rangle\bigg\langle\parfig{.04}{3body_2v3}\bigg|.
\end{align}
Furthermore, $V_{2,\vec{r}}$ and $V_{3,\vec{r}}$ are $3\times 3$ and $2\times 2$ Haar random unitaries respectively. Intuitively, $U_{C,\vec{r}}$ updates collisions between particles only for the corresponding two body and three body configurations shown in Fig.~\ref{fig:fig1}. We note that the $U_{C,\vec{r}}$, commute with one another since: (1) degrees of freedom with arrows pointing outward from a site $\vec{r}$ are not shared with another site; (2) a collision gate at site $\vec{r}$ acts as the identity on degrees of freedom with arrows pointing outward on a different site $\vec{r}\,'$.  \par
The propagation step is a deterministic step that is enacted by the unitary operator $\mathcal{U}_P = \prod_{\ell=1}^{3} U_{P,\ell}^{\text{inter}} U_{P,\ell}^{\text{intra}}$ where $U_{P,\ell}^{\text{inter}}$ and $U_{P,\ell}^{\text{intra}}$ are given by


\begin{align}
U_{P,\ell}^{\text{intra}} &= \prod_{\bar{\vec{x}}} \text{SWAP}_{(\vec{x}-\hat{b}_\ell,\ell+3),(\vec{x},\ell)},\\
U_{P,\ell}^{\text{inter}} &= \prod_{\bar{\vec{x}}} \text{SWAP}_{(\vec{x},i),(\vec{x},\ell+3)},
\end{align}
where $\text{SWAP}_{(\vec{x},\ell),(\vec{y},\ell')}$ denotes the usual SWAP operator which interchanges degrees of freedom on link $\ell$ associated with $\vec{x}$ and with degrees of freedom on link $\ell'$ associated with $\vec{y}$. The unit vector $\hat{b}_\ell$ is given by $\hat{b}_{\ell} = (\cos(\frac{2\pi \ell}{6}),\sin(\frac{2\pi \ell}{6}))$ and $\ell+3$ implicitly should be evaluated modulo six.\par

%
%
From the expression one can see that $\mathcal{U}_P$ is a brickwork circuit of SWAP gates acting on the three different reflection axes of the hexagonal cell. The first layer of the SWAP circuit swaps particles residing on the same link and the second layer of the SWAP circuit swaps particles associated to the same site but on different nearest neighbor links (along one of the reflection axes of the hexagonal cell) as shown in Fig.~\ref{fig:fig1}. This procedure ensures that particles will move in the direction their arrow points to a link one lattice spacing away.\par
\emph{Hydrodynamics}.---With the dynamics of the model specified, we now identify the conservation laws of the model and determine the hydrodynamics up to the diffusive scale of an individual circuit. Any given circuit realization conserves the following operators:
\begin{align}
N&=\sum_{\vec{x}} \mathfrak{n}_{\vec{x}}= \sum_{\vec{x}}\sum_{\ell=1}^{6} n_{\vec{x}}^{\ell},\\
\vec{P} &= \sum_{\vec{x}} \vec{\mathfrak{p}}_{\vec{x}} =\sum_{\vec{x}}\sum_{\ell=1}^{6}\hat{b}_{\ell} n_{\vec{x}}^{\ell},
\end{align}
 where $n_{\vec{x}}^{\ell}$ corresponds to the occupation number on the link $\ell$ of site $\vec{x}$. The first conservation law reflects total particle number conservation while the second reflects a type of ``momentum" conservation---although this operator is not associated with the generator of translations. In this context, ``momentum" conservation refers to the fact that the vector sum of the particles' direction of motion is conserved.  One can see that this is indeed the case since collision unitary gates only cause transitions between states whose vector sum is zero and the propagation step does not change the direction in which particles travel. 
 \par

The momentum operator $ \vec{\mathfrak{p}}_{\vec{x}} $ also coincides exactly with the particle current: this current cannot relax, corresponding to dissipationless, ballistic particle transport. 
Denoting the expectation values of the local conserved quantities $n(\vec{x},t) = \langle \mathfrak{n}_{\vec{x}} \rangle$ and $n(\vec{x},t)\vec{v}(\vec{x},t) =  \langle \mathfrak{p}_{\vec{x}} \rangle
$, the long time dynamics of a typical FHP circuit are expected to be governed by the following hydrodynamics equations
\begin{align}
\quad\partial_{t}n + \partial_{i}(n v_{i}) &=0, \notag \\
\partial_{t}(n v_{i}) + \partial_{j} \langle  \pi_{ij} \rangle &= 0.
\label{eqn:singlehydro}
\end{align}
The first equation corresponds to momentum conservation, while the expectation value of the stress tensor $\pi_{ij}(\vec{x}) = \sum_{\ell=1}^6 \hat{b}_{\ell}^i\hat{b}_{\ell}^j n_{\vec{x}}^{\ell}$ can be expressed as a gradient expansion up to the diffusive scale $ \langle \pi_{ij} \rangle = P \delta_{ij} + A_{ijkl} c^2 ng(n)v_{k}v_{l} - (\eta-\tilde{\eta})A_{ijkl}\partial_k (n v_l) $,
with $A_{ijkl} = \delta_{ik}\delta_{jl}+\delta_{il}\delta_{jk}-\delta_{ij}\delta_{kl}$, $P = c^2 n$, $c=1/\sqrt{2}$, $\tilde{\eta} = 1/8$ and $g(n) = \frac{3-n}{6-n}$~\cite{suppmat}.\par
These are not quite the Navier-Stokes equations due to effects from the lattice: (1) the factor $g(n)$ which breaks Galilean invariance, (2) the tensor $A_{ijkl}$ signifying that the dynamics is invariant under the group of symmetries of a regular hexagon. The latter feature has been noted in other works exploring hydrodynamics of hexagonal Fermi surfaces~\cite{Cook:2019aa}. We note that the $A_{ijkl}$ is invariant under the full rotation group~\cite{lgahydro1990} and so the above equations still describe an isotropic fluid. Additionally, one can show that when the average density is approximately constant, i.e.~$n\simeq n_0$, one recovers the incompressible Navier-Stokes equations~\cite{lgahydro1990}.\par
To characterize the hydrodynamics of an individual circuit we will examine linear response coefficients. For the above hydrodynamics one finds two damped sounds modes and one diffusive mode characterized by the speed of sound, $c$, and shear viscosity $\eta$. In the next section, we will study sample-to-sample fluctuations by studying fluctuations of these transport coefficients. In particular, we will show that the speed of sound is fixed to the same value for every circuit realization. \par
\begin{figure}[!t]
	\centering
	\includegraphics[width=\columnwidth]{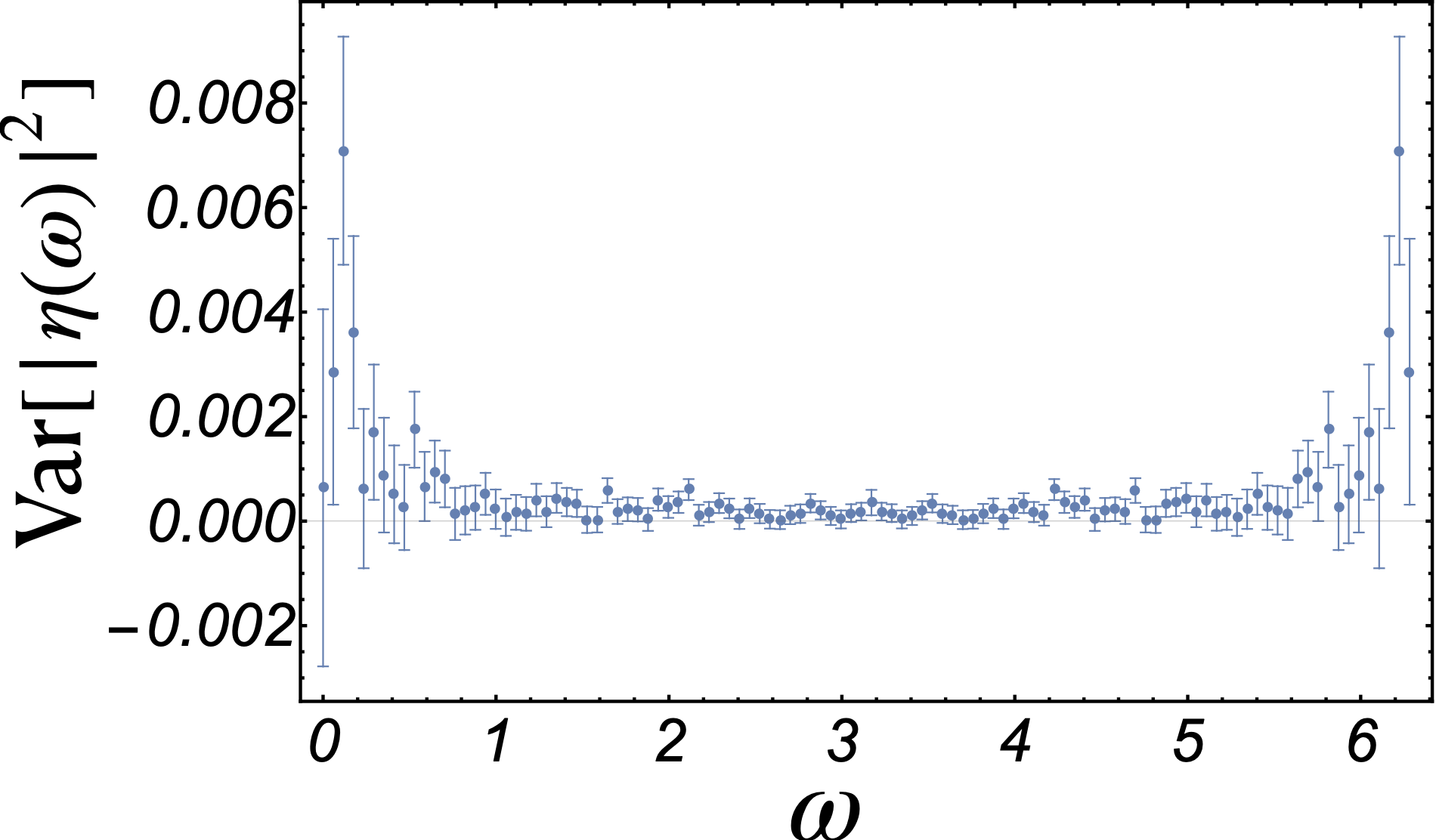}
	\caption{{\bf Sample-to-sample fluctuations of shear viscosity.} The fluctuations of the a.c.~shear viscosity evaluated using the effective Markov process from the large-$d$ limit of the statistical mechanics model. This was evaluated for a system with $216\times216$ sites at a time $T=108$. Data is averaged over $8.9 \times 10^6$ realizations and $\mu=0.5$.}
	\label{fig:fig2}
\end{figure}
\emph{Fluctuations in transport coefficients}.--- As mentioned in the previous section there are two linear response coefficients which we can diagonose transport with. We begin by discussing the fluctuations in the speed of sound, $c$. Owing to the conservation of ``momentum", $\vec{P}$, sound modes appear at the Euler scale for hydrodynamics. Since all transport coefficients at the Euler scale are fixed by thermodynamics~\cite{10.21468/SciPostPhysLectNotes.18}, so is the speed of sound, in particular $c=\sqrt{\frac{\partial P}{\partial n}}$. 
Thus the speed of sound only depends on thermodynamic expectation values, and does not depend on details of the random Haar gates: it is fixed to $c=1/\sqrt{2}$ for all circuit realizations.\par 
For fluctuations of the shear viscosity, we know that the associated current, i.e. the momentum stress tensor, is not conserved and so the shear viscosity is generally not set by thermodynamics so we must compute fluctuations from circuit to circuit via other means. In general sample-to-sample fluctuations are generically encoded in non-linear observables, i.e.~observables which are non-linear functions of the density matrix. In this work we will compute the variance from circuit to circuit of the a.c.~shear viscosity, $\text{Var}[\eta(\omega)]$. For each circuit realization, the a.c.~shear viscosity can be related to autocorrelation functions of the ``momentum" stress tensor via the Kubo formula~\cite{kubo1957statistical},
\begin{equation}
\eta(\omega) = \frac{1}{2} C(0) + \sum_{t=1}^{\infty} e^{i\omega t} C(t),
\end{equation}
where  $C(t) = \frac{1}{V} \langle \Pi_{xy}(t) \Pi_{xy}(0) \rangle$ with $V$ the volume of the system and $\Pi_{ij} = \sum_{\vec{x}} \pi_{ij}(\vec{x})$.  In this case $\langle A \rangle \equiv \frac{1}{Z}\text{tr}(A e^{-\mu N})$ with $\mu$ denoting the chemical potential.\par
Evaluating the average of this quantity over circuits involves computing the Haar average of $\mathcal{U} \otimes \mathcal{U}^{*}$, whereas computing the variance over circuit realizations requires averaging the two-copy replicated unitary, $(\mathcal{U} \otimes \mathcal{U}^{*})^{\otimes 2}$. Using standard tools developed for one-dimension quantum circuits~\cite{Fisher_2023, Agrawal:2022aa,Nahum:2018aa,Zhou:2019aa,Vasseur:2019aa,Jian:2020aa,Bao:2020aa,Li:2021aa}, we find that this averaging procedure maps the single copy Haar average onto a classical irreversible Markov process, while the two copy average maps onto a statistical mechanics model with permutation degrees of freedom $\sigma \in S_2$ corresponding pairings of the replica living on the vertices and charge degrees of freedom on the edges~\cite{suppmat}.\par
 In the limit of infinite ancilla dimension, $d\to\infty$, 
 the average of the two-copy replica unitary corresponds to two decoupled copies of $\mathbb{E}[\mathcal{U} \otimes \mathcal{U}^{*}]$, i.e. two decoupled classical Markov processes~\cite{McCulloch:2023aa,suppmat}. Generalizing the one dimensional results of Ref.~\cite{McCulloch:2023aa}, we find that at large but finite $d$, one can systematically account for corrections and consequently show that leading order correction corresponds to a new effective Markov process which couples the two single-copy Markov processes~\cite{suppmat}.\par
This allows us to study the variance of the viscosity over quantum circuits through efficient numerical simulations---presented in Fig.~\ref{fig:fig2} . In Fig.~\ref{fig:fig2}, the variance of the a.c. shear viscosity is computed at a finite time $T = L/2$ where $L$ is the linear size of the system with the ancilla dimension, $d=2$. We observe that the variance is consistent with vanishing for the given amount of sampling we were able to achieve. 
 %
Since the variance of the a.c. shear viscosity appears to show vanishing fluctuations, this indicates that a typical circuit will also be characterized by the a.c.~shear viscosity of the ensemble average with corrections that are vanishingly small. Given that the speed of sound and a.c.~shear viscosity correspond to their ensemble averaged values, we can now study the hydrodynamics of a typical circuit via its ensemble averaged dynamics.\par  
\emph{Hydrodynamic transport}.---We illustrate this mapping by computing the unequal time density correlation functions as well as momentum density correlation functions of typical circuits by evaluating their ensemble average.  These quantities are linear functions of the density matrix, since they are of the form $\text{tr}(O(t)O'(0)\rho)$, and hence the ensemble averaged dynamics can be described by a classical Markov process~\cite{Khemani:2018aa,Rakovszky:2018aa,Friedman:2019aa,Singh:2021aa,Moudgalya:2021aa}. More precisely,
\begin{equation}
\mathbb{E}[\text{tr}(O(t)O'(0)\rho)] = \text{tr}(\mathcal{T}^{t}[O(0)]O'(0)\rho)
\end{equation}
where $\mathcal{T}$ is a superoperator (transfer matrix), which is positive semi-definite and is Markovian, i.e. $\sum_{i} \mathcal{T}_{ij}  = 1$. Importantly, this Markov process is irreversible because of the propagation step of the model. 
\par
\begin{figure}[!t]
	\centering
	\includegraphics[width=\columnwidth]{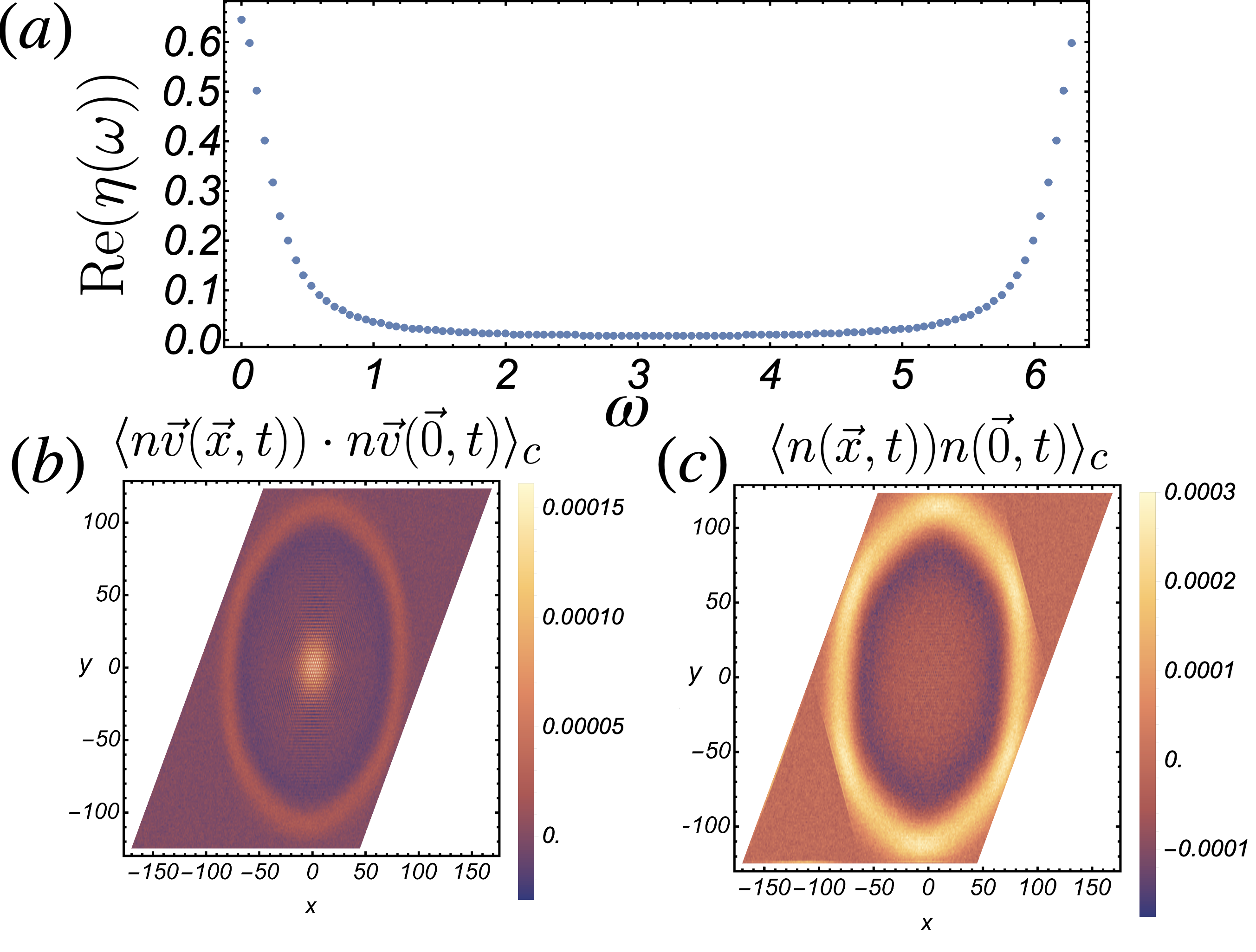}
	\caption{{\bf Ensemble averaged structure factors and shear viscosity.} (a) The real part of the a.c.~shear viscosity. (b) Momentum density structure factor features a ballistic portion corresponding to the sound modes and a heat mode located at the center. (c) One sees that the particle density structure factor also features a clear ballistic behavior indicative of sound modes. The above data was evaluated at $T=108$ for a system size of $216\times216$ and the data is averaged over $8.9 \times 10^6$ realizations and $\mu=0.5$.}
	\label{fig:fig3}
\end{figure}
For the FHP random unitary circuit, one can show the corresponding Markov process is a variant of the well-known FHP lattice gas automaton~\cite{Frisch:1986aa,lgahydro1990}. Like the random unitary circuit, the transfer matrix of this FHP variant generates stochastic dynamics consisting of a random collision step, denoted by $\mathcal{T}_{C}$, followed by a deterministic propagation step in which particles move one lattice spacing in which their arrow points, $\mathcal{T}_P$. So the full transfer matrix is given by $\mathcal{T}=\mathcal{T}_P\mathcal{T}_C$. 
Like the random unitary circuit, the collision portion has a gate structure, i.e. $\mathcal{T}_C = \prod_{\vec{x}} T_{C,\vec{x}}$. The collision gates for the model are then given by
\begin{equation}
T_{C,\vec{r}} = \frac{1}{3}\mathcal{P}_{2}+\frac{1}{2}\mathcal{P}_{3}+\mathcal{P}_\perp,
\end{equation}
where $\mathcal{P}_{2}$ projects onto the symmetrized state $\frac{1}{3}\left(\big|\parfig{.035}{2body_1v3}\big) + \big|\parfig{.035}{2body_2v3}\big) + \big|\parfig{.035}{2body_3v3}\big)\right)$, $\mathcal{P}_{3}$ projects onto the state $\frac{1}{2}\left(\big|\parfig{.035}{3body_1v3}\big) + \big|\parfig{.035}{3body_2v3}\big)\right)$ and $\mathcal{P}_{\perp}$ projects onto the space orthogonal to the two and three-particle collision spaces (here we have denoted the classical states for a particle configuration $\Gamma$ by $|\Gamma)$). This model has equiprobable transition to any state within a given collision subspace and all other states are unchanged. We note that the present stochastic evolution differs from the original FHP model since collision updates have a probability to not reconfigure particles in the two body or three body collision subspaces---such a modification will only change transport coefficients but not the hydrodynamic behavior.\par
Using the transfer matrix, we numerically obtain the structure factors for the particle and ``momentum" density, $\langle n(\vec{x},t)n(\vec{0},0)\rangle_{c}$ and $\langle (n\vec{v})(\vec{x},t)\cdot (n\vec{v})((\vec{0},0)\rangle_{c}$ as well as real part of the a.c.~shear viscosity. Our results are shown in Fig.~\ref{fig:fig3} and we indeed observe the presence of two damped sound modes and a heat mode, as expected from standard linearized hydrodynamics. \par
 \emph{Discussion}.--- 
In this letter, we constructed an ensemble of random quantum circuit whose hydrodynamics corresponds to the incompressible Navier-Stokes equations. We characterized sample-to-sample fluctuations of transport coefficients such as the a.c.~shear viscosity by mapping fluctuations onto an effective classical statistical mechanics model. Our results are consistent with fluctuations vanishing at sufficiently long times, indicating that a typical circuit behaves as its ensemble average and furthermore a typical circuit has a shear viscosity that is effectively the same as the ensemble average.\par
Since sample-to-sample fluctuations of transport coefficients are quantitatively very small, we are able to study the hydrodynamics of typical circuit via its  ensemble averaged dynamics. We show that the ensemble averaged dynamics corresponds to a variant of a well-known classical Markov process known as the FHP model, and numerically verify that the system does host two sound modes and a heat mode. The presence of sound modes places these random circuits as a rare example of a lattice non-integrable system with---albeit fine-tuned---ballistic transport and an exact Drude weight. \par
%
Strikingly, the hydrodynamics of the coupled model features an additional conservation law coming from the presence of anti-commuting charges in the dynamics~\cite{McNamara:1988aa}. Although the effects of the anticommuting charges do not appear to affect typical thermal states~\cite{Kadanoff:1989aa}, it would be interesting to see how the presence of these new conservation laws changes the approach towards the ensemble average.\par
We also remark that the Navier-Stokes hydrodynamics in two dimensions is not stable to stochastic noise which results in logarithmic corrections to the hydrodynamics~\cite{Forster:1977aa}. This has been observed in the original FHP model by studying the finite size scaling of the d.c.~shear viscosity~\cite{Kadanoff:1989aa}. It would be an interesting future work to see if such effects have any consequence on quantum fluctuations.

\par
\emph{Acknowledgements}.--- We thank Ethan Lake for stimulating discussions, and Thomas Scaffidi for pointing out the similarities to hydrodynamics of polygonal Fermi surfaces. This work was supported by the  US Department of Energy, Office of Science, Basic Energy Sciences, under award No. DE-SC0023999 (H.S. and R.V.) and through the Co-design Center
for Quantum Advantage (C2QA) under contract number DE-SC0012704 (S.G.).
\bibliography{lgarefs}

\bigskip
\onecolumngrid
\newpage


\section*{Supplementary material}

# Supplemental Material for "Emergence of Navier-Stokes hydrodynamics in chaotic quantum circuits"

Hansveer Singh[1], Ewan McCulloch[2], Sarang Gopalakrishnan[2] and Romain Vasseur[1]

[1]*Department of Physics, University of Massachusetts,*
*Amherst, Massachusetts 01003, USA*
[2]*Department of Electrical and Computer Engineering,*
*Princeton University, Princeton NJ 08544, USA*

## I. STATISTICAL MECHANICS MODEL

In this section we show how to map the variance of correlation functions, $C(t) \equiv \text{tr}(O(t)O(0)\rho(0))$ onto a statistical mechanics model where $O$ is an operator and $\rho$ is the state of our system. The variance of $C(t)$ is given by $\text{Var}[C(t)] = \mathbb{E}[C(t)^2] - \mathbb{E}[C(t)]^2$, where $\mathbb{E}[\cdot]$ denotes averaging over the Haar ensemble of unitary operators corresponding to the collision gates. When evaluating such averages it is convenient to vectorize operators, i.e. we use the Choi map $O = \sum_{ij} O_{ij}|i\rangle\langle j| \to |O\rangle = \sum_{ij} O_{ij}|ij\rangle$ where $|i\rangle$ is a complete basis over the Hilbert space, $\mathcal{H}$. The vectorized Hilbert space of operators corresponds to $\mathcal{H} \otimes \mathcal{H}^*$ with the following inner product: $\langle A|B\rangle = \text{tr}(A^\dagger B)$ which corresponds to the usual Hilbert-Schmidt norm. Upon vectorization,

$$C(t) = \langle O|\mathcal{U}^t \otimes \mathcal{U}^{*t}|O\rho\rangle, \tag{1}$$

where $\mathcal{U}$ is the unitary evolution operator for a single time step.

Additionally it becomes useful to evaluate $C(t)^2$ over an enlarged—so-called replica—Hilbert space for states given by $\mathcal{H} \otimes \mathcal{H}$. Explicitly one can write $C(t)^2 = \text{tr}_{\mathcal{H} \otimes \mathcal{H}}(O(t)O(0)\rho(0) \otimes O(t)O(0)\rho(0))$ where the first (second) $O(t)O(0)\rho(0)$ belongs to the first (second) tensor factor of the replicated Hilbert space. Vectorizing the expression for $C(t)^2$ in the replicated Hilbert space (which corresponds to $\mathcal{H} \otimes \mathcal{H}^* \otimes \mathcal{H} \otimes \mathcal{H}^*$) one obtains,

$$C(t)^2 = \langle O \otimes O|\mathcal{U}^t \otimes \mathcal{U}^{*t} \otimes \mathcal{U}^t \otimes \mathcal{U}^{*t}|O\rho \otimes O\rho\rangle. \tag{2}$$

To evaluate Eq. 1 and Eq. 2 we must compute $\mathbb{E}[\mathcal{U}^t \otimes \mathcal{U}^{*t}]$ and $\mathbb{E}[\mathcal{U}^t \otimes \mathcal{U}^{*t} \otimes \mathcal{U}^t \otimes \mathcal{U}^{*t}]$ which are typically referred to as the single copy and two copy average, respectively. In the next subsections we first evaluate the two copy average and then the single copy average in the large-$d$ limit (recall $d$ corresponds to the dimension of the ancilla degrees of freedom in the main text).

### A. Effective Markov Process

Our central result will be that the two copy average corresponds to a statistical mechanics model which reduces to an effective Markov process in the large-$d$ limit. To arrive at this result we note that the unitary evolution only contains Haar random unitaries for the collision step and collision gates are identical independently distributed Haar random unitaries at each position and time step. Thus, we only need to evaluate $\mathbb{E}[U_C \otimes U_C^* \otimes U_C \otimes U_C^*]$ where $U_C$ corresponds to the collision gate in the main text. We closely follow the calculation done in Ref [1] for $U(1)$ charge conserving random unitary unitary circuits. In particular, we will associate to each qubit (which either hosts a particle or is vacant) an ancilla qudit. The ancilla follows the same updates in the propagation step of the evolution as the qubits they accompany.

#### 1. Two Copy Haar Average

Since our Haar random collision gates trivially rotate all the particle configurations except for the three two-particle collisions and the two three-particle collision, we will find it useful to separate out these two subspaces in our nomenclature. Let us refer to the subspace of two-particles collisions and three-particles collisions as $\mathcal{H}_2 \subset \mathcal{H}_\bigcirc$ and $\mathcal{H}_3 \subset \mathcal{H}_\bigcirc$ where $\mathcal{H}_\bigcirc$ is the full Hilbert space of states acted on by the collision gate. The remaining particle configurations are not mixed by the unitary (although the ancilla does receive a random rotation), and so each is a conserved sector. We refer to each of these conserved sectors as $\mathcal{H}_{\perp_k}$, where $k$ enumerates each particle configuration.

We have used the $\perp$ symbol to indicate that the subspaces $\mathcal{H}_{\perp_k}$ are all orthogonal to the two-particle and three-particle collision sectors $\mathcal{H}_2$ and $\mathcal{H}_3$.

With this set up, we can define the unitary gate as $U_C = \sum_{a \in 2,3,\{\perp_k\}} W_a$, with $W_a = P_a V_a P_a$. $P_2$ and $P_3$ correspond to the two body and three body collision subspaces outlined in the main text and $P_{\perp_k}$ projects onto the conserved sector $\mathcal{H}_{\perp_k}$. The matrices $V_2$ and $V_3$ are identical independently distributed $3d^6 \times 3d^6$ and $2d^6 \times 2d^6$ Haar random unitaries, respectively, whereas $V_{\perp_k}$ is a $d^6 \times d^6$ Haar random unitary acting only on the ancilla degrees of freedom.

The two copy average is given by,

$$\mathbb{E}[U_C \otimes U_C^* \otimes U_C \otimes U_C^*] = \sum_{\substack{a,b \in 2,3,\{\perp_k\} \\ a \neq b}} \left( \mathbb{E}[W_a \otimes W_a^* \otimes W_b \otimes W_b^*] + \mathbb{E}[W_a \otimes W_b^* \otimes W_b \otimes W_a^*] \right) \\ + \sum_{a \in 2,3,\{\perp_k\}} \mathbb{E}[W_a \otimes W_a^* \otimes W_a \otimes W_a^*], \qquad (3)$$

where we have only written down terms that give non-zero contributions after Haar averaging.

Define the following state for the $\ell$-th link of the six links acted on my the collision unitary

$$\left| n_1^{(l)}, n_2^{(l)}; \sigma \right\rangle = \sum_{\alpha_1 \in \mathcal{H}_{n_1^{(l)}}} \sum_{\alpha_2 \in \mathcal{H}_{n_2^{(l)}}} \left| \alpha_1, \alpha_{\sigma(1)}^*, \alpha_2, \alpha_{\sigma(2)}^* \right\rangle, \qquad (4)$$

where $\sigma \in S_2$ is either the identity permutation or the swap permutation (i.e., a transposition), and where $\mathcal{H}_{n^{(\ell)}}$ is the subspace of states on link $\ell$ which have a particle occupancy $n^{(\ell)}$. Since $n = 0, 1$ this subspace is just the size of the ancilla. For all six links, we can similarly define the following

$$|\boldsymbol{n}_1, \boldsymbol{n}_2; \sigma\rangle = \left( \bigotimes_{\text{links } \ell} \left| n_1^{(\ell)}, n_2^{(\ell)}; \sigma \right\rangle \right). \qquad (5)$$

where the $\boldsymbol{n}_i = \{n_i^{(\ell)}\}_{\ell=1}^6$ is a vector of the occupations on replica $i$ for each of the six links. We will find it useful to define one more state using the previous definitions,

$$|a, b; \sigma\rangle = \sum_{\substack{\boldsymbol{n}_1 \in a \\ \boldsymbol{n}_2 \in b}} |\boldsymbol{n}_1, \boldsymbol{n}_2; \sigma\rangle, \qquad (6)$$

where $a, b \in 2, 3, \{\perp_k\}$ and where all links have the same permutation $\sigma$. We have also used the notation $\boldsymbol{n} \in a$ to indicate that we are summing over states which particle configurations in the subspace $\mathcal{H}_a$.

Equipped with these definitions, we can write the two-copy Haar average as

$$\mathbb{E}[U_C \otimes U_C^* \otimes U_C \otimes U_C^*] = \sum_{\substack{a,b \in 2,3,\{\perp_k\} \\ a \neq b}} \frac{1}{d_a d_b} \left[ |a, b; \mathbb{1}\rangle \langle a, b; \mathbb{1}| + |a, b; \text{Sw}\rangle \langle a, b; \text{Sw}| \right] \\ + \sum_{a \in 2,3,\{\perp_k\}} \left[ \text{Wg}_{d_a}(\mathbb{1}) \left( |a, a; \mathbb{1}\rangle \langle a, a; \mathbb{1}| + |a, a; \text{Sw}\rangle \langle a, a; \text{Sw}| \right) \right. \\ \left. + \text{Wg}_{d_a}(\text{Sw}) \left( |a, a; \mathbb{1}\rangle \langle a, a; \text{Sw}| + |a, a; \text{Sw}\rangle \langle a, a; \mathbb{1}| \right) \right], \qquad (7)$$

where where $d_2 = 3d^6$, $d_3 = 2d^6$, and $d_a = d^6$ otherwise, and where $\text{Wg}_{d_a}(\sigma)$ are the Weingarten functions for $\sigma \in S_2$. Specifically, we have $\text{Wg}_{d_a}(\mathbb{1}) = \frac{1}{d_a^2 - 1}$ and $\text{Wg}_{d_a}(\text{Sw}) = -\frac{1}{d_a(d_a^2 - 1)}$. We have denoted the identity permutation as $\mathbb{1}$ and the swap permutation as Sw.

In terms of the link states $|\boldsymbol{n}_1, \boldsymbol{n}_2; \sigma\rangle$ defined in Eq. 5, Eq. 7 becomes

$$\mathbb{E}[U_C \otimes U_C^* \otimes U_C \otimes U_C^*] = \sum_{\boldsymbol{n}_1, \boldsymbol{n}_2} \sum_{\boldsymbol{n}_1', \boldsymbol{n}_2'} \sum_{\sigma, \tau} T^{(2)}_{(\boldsymbol{n}_1, \boldsymbol{n}_2; \sigma), (\boldsymbol{n}_1', \boldsymbol{n}_2'; \tau)} |\boldsymbol{n}_1, \boldsymbol{n}_2; \sigma\rangle \langle \boldsymbol{n}_1', \boldsymbol{n}_2'; \tau|, \qquad (8)$$

where the tensor $T^{(2)}$ is non-zero only for states for which $\boldsymbol{n}_1$ ($\boldsymbol{n}_2$) and $\boldsymbol{n}_1'$ ($\boldsymbol{n}_2'$) are in the same sector $a$ ($b$). Assuming

this to be true, then we have

$$T^{(2)}_{(\boldsymbol{n}_1,\boldsymbol{n}_2;\sigma),(\boldsymbol{n}'_1,\boldsymbol{n}'_2;\tau)} = \begin{cases} (d_a d_b)^{-1}\delta_{\sigma,\tau} & \text{if } a \neq b \\ d_a^{-2-|\sigma\tau^{-1}|} & \text{if } a = b \in \{\perp_k\} \\ \mathrm{Wg}_{d_a}(\sigma\tau^{-1})\left[\frac{d_a^{|\sigma\tau^{-1}|}}{d^6}(\delta_{\boldsymbol{n}_1,\boldsymbol{n}_2}+\delta_{\boldsymbol{n}'_1,\boldsymbol{n}'_2}) + \frac{d_a}{d_a^{|\sigma\tau^{-1}|}}\left(1 + \frac{1}{d^{12}}\delta_{\boldsymbol{n}_1,\boldsymbol{n}_2}\delta_{\boldsymbol{n}'_1,\boldsymbol{n}'_2}\right)\right] & \text{if } a = b \in 2,3, \end{cases} \tag{9}$$

where $|\sigma|$ is the transposition distance of $\sigma$ from the identity permutation.

We can now interpret the evolution induced by the averaged two-copy gates, followed by a propagation step, as a statistical mechanics model with charge degrees of freedom $(\mathbf{n}_1, \mathbf{n}_2)$ residing on the edges and permutation degrees of freedom residing on the vertices as shown in Fig. 1. The matrix elements of $T^{(2)}$ will be interpreted as Boltzmann weights for the internal edges of the collision gates (the thick edges in Fig. 1).

So far we have restricted our attention to a single collision gate. The full unitary evolution for a single time step, $\mathcal{U}$, also has a propagation step, $\mathcal{U}_P$. $\mathcal{U}_P$ propagates degrees of freedom along the link directions $\hat{b}_i$ away from the center of the hexagonal cell to the same link in the next hexagonal cell (as shown in Fig. 1 in the main text). The circuit of SWAP gates described in the main text that accomplishes the above propagation step is equivalent to contracting the indices of each link $\ell$ of a collision gate at $(\vec{x},t)$ with the indices of the collision gates at locations $(\vec{x}+\hat{b}_\ell,t+1)$. This contraction is shown in Fig. 1(b) and give rise to a Boltzmann weight in the statistical mechanics model proportional to the overlaps $\langle n_1^{(\ell)}, n_2^{(\ell)}, \sigma | n_1^{(\ell)}, n_2^{(\ell)}, \tau \rangle$, where $\sigma$ and $\tau$ are the permutations associated to a vertex at $(\vec{x},t)$ and $(\vec{x}+\hat{b}_\ell,t+1)$ respectively (the charge label must be the same in the two contracted states as it is a property of the edge). The Boltzmann weights are given by

$$\frac{1}{d^2}\left\langle n_1^{(\ell)}, n_2^{(\ell)}; \sigma \middle| n_1^{(\ell)}, n_2^{(\ell)}; \tau \right\rangle = \delta_{n_1^{(\ell)}, n_2^{(\ell)}} d^{-|\sigma\tau^{-1}|} + (1-\delta_{n_1^{(\ell)}, n_2^{(\ell)}})\delta_{\sigma,\tau}. \tag{10}$$

The Boltzmann weights for the edges internal to a collision gate, and those for the edges connecting collision gates are summarized in Fig. 1(a),(c).

The discussion so far has summarized the 2+1d statistical mechanics model associated with $\mathbb{E}[\mathcal{U}^t \otimes \mathcal{U}^{*t} \otimes \mathcal{U}^t \otimes \mathcal{U}^{*t}]$. This quantity will correspond to the average of $C(t)^2$ once we impose boundary conditions at the initial and final steps of the statistical mechanics model we just constructed. In the evaluation of $C(t)^2$ we are evaluating a matrix element between the states $|O\rho \otimes O\rho\rangle$ and $|O \otimes O\rangle$. Since these states factorize over the replicas, i.e., $|O\rho \otimes O\rho\rangle = |O\rho\rangle \otimes |O\rho\rangle$, these states impose the permutation boundary conditions $\sigma_{\text{initial}} = \mathbb{1}$ and $\sigma_{\text{final}} = \mathbb{1}$.

In the next sections we will take the large-$d$ limit and demonstrate the statistical mechanics model we just constructed maps onto an effective Markov process.

### 2. Infinite d limit and single copy average

The Boltzmann weights for the thick edges is $\mathcal{O}(d^{-12})$ if the two permutations are equal $\sigma = \tau$, and $\mathcal{O}(d^{-18})$ if $\sigma \neq \tau$. Similarly, for the edges connecting collision gates, the Boltzmann weights are $\mathcal{O}(1)$ if the two permutations are equal $\sigma = \tau$ and $\mathcal{O}(d^{-1})$ if $\sigma \neq \tau$. Therefore, in the infinite $d$ limit, all configurations in which a pair of neighbouring permutations are unequal have zero contribution to the partition sum. Because of the connectivity of the vertices of the two-copy statistical mechanics model, and because of the boundary conditions, infinite $d$ simply sets every permutation to be equal to the identity permutation, $\sigma = \mathbb{1}$. The partition sum being dominated by a single orientation of permutations reflects the fact that replica statistical mechanics models have ferromagnetic ground states [2–5].

Setting $\sigma = \mathbb{1}$ at every vertex in the statistical mechanics model is equivalent to keeping only the terms in the two-copy Haar average in Eq. 7 which have identity permutations. Doing this yields

$$\mathbb{E}[U_C \otimes U_C^* \otimes U_C \otimes U_C^*] \to \sum_{\substack{a,b \in 2,3,\{\perp_k\} \\ a \neq b}} \frac{1}{d_a d_b} |a,b;\mathbb{1}\rangle \langle a,b;\mathbb{1}| + \sum_{a \in 2,3,\{\perp_k\}} \mathrm{Wg}_{d_a}(\mathbb{1}) |a,a;\mathbb{1}\rangle \langle a,a;\mathbb{1}|, \tag{11}$$

$$\approx \sum_{a,b \in 2,3,\{\perp_k\}} \frac{1}{d_a d_b} |a,b;\mathbb{1}\rangle \langle a,b;\mathbb{1}|, \tag{12}$$

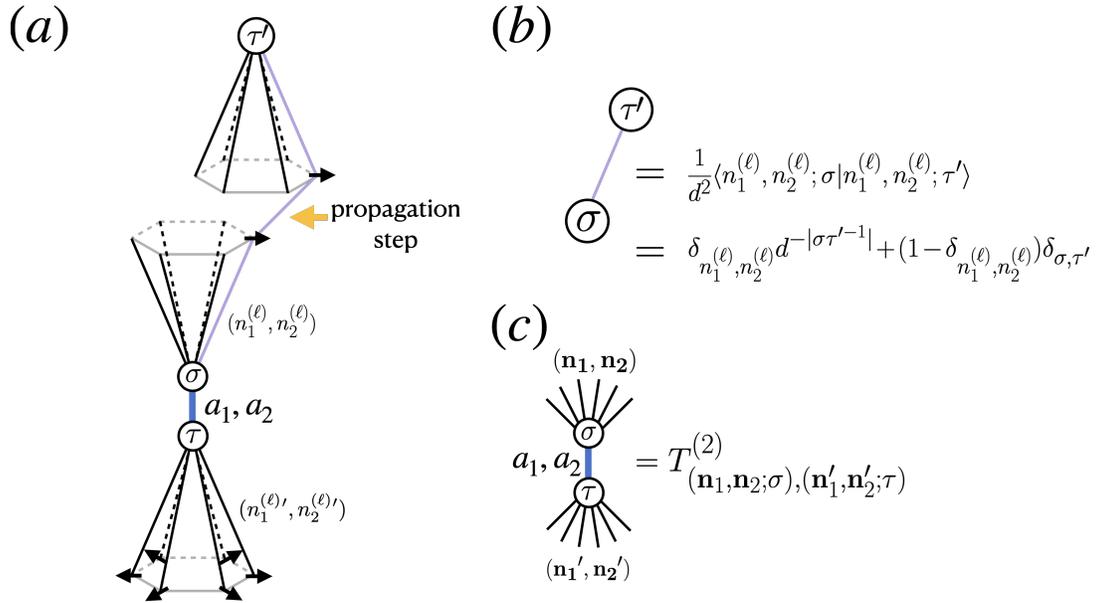

FIG. 1. **Two Copy Statistical Mechanics Model.** (a) Charge degrees of freedom live on links and permutation degrees of freedom are located on sites. The arrows indicate the direction any particles living on links in the triangular lattice will be propagated in during the propagation step. (b) The weights associated with contracting links between different collision step transfer matrices. (c) Graphical representation of the collision step transfer matrix $T^{(2)}$, whose matrix elements are Boltzmann weights in the statistical mechanics model.

where to get to the second line we used the fact that we are taking $d \to \infty$, to make the replacement $\text{Wg}_{d_a}(\mathbb{1}) \to d_a^{-2}$. A short calculation gives us the one-copy average,

$$\mathbb{E}[U_C \otimes U_C^*] = \sum_{a \in 2,3,\{\perp_k\}} \frac{1}{d_a} |P_a\rangle \langle P_a|, \quad |P_a\rangle \equiv \sum_{\alpha \in \mathcal{H}_a} |\alpha, \alpha^*\rangle. \tag{13}$$

Comparing to Eq. 11 and using the definitions in Eq. 4 and 5, we see that in the infinite $d$ limit (and with the $\sigma = \mathbb{1}$ boundary conditions), the two-copy Haar averaged model simply reduces to the tensor product of two single-copy averaged models,

$$\mathbb{E}[U_C \otimes U_C^* \otimes U_C \otimes U_C^*] \to \mathbb{E}[U_C \otimes U_C^*] \otimes \mathbb{E}[U_C \otimes U_C^*]. \tag{14}$$

A consequence of this result is that the variance in the $d \to \infty$ limit vanishes to leading order. We also note that in the single-copy average, the ancilla factorizes out,

$$\mathbb{E}[U_C \otimes U_C^*] = \left(\frac{1}{3}|P_2\rangle\langle P_2| + \frac{1}{2}|P_3\rangle\langle P_3| + P_\perp\right)_{\text{particles}} \otimes \left(\frac{1}{d^6}|\mathbb{1}\rangle\langle\mathbb{1}|\right)_{\text{ancilla}}, \quad P_\perp \equiv \sum_{a \in \{\perp_k\}} |P_a\rangle\langle P_a|. \tag{15}$$

The term before the tensor product acts only on the particle degrees of freedom and the term after the tensor product acts only on the ancilla qudits.

Since we are interested only in the particle dynamics, we will marginalize over the ancilla degrees of freedom at the final time, this is equivalent to contracting each ancilla with the identity matrix $|\mathbb{1}\rangle$. As we can see from Eq. 15, this state can just be propagated through to the initial state, where we find $\langle \rho_{\text{ancilla}} | \mathbb{1} \rangle = 1$. I.e,., the ancilla can be dropped entirely and we are left with a trasnfer matrix for the particle degrees of freedom only, given by $\frac{1}{3}|P_2\rangle\langle P_2| + \frac{1}{2}|P_3\rangle\langle P_3| + P_\perp$. This is just the transfer matrix of a Markov process with random two and three body collisions with associated probabilities $1/3$ and $1/2$, respectively. Similarly the two copy average reduces to two decoupled copies of this classical Markov process for the charge dynamics to leading order in the $d \to \infty$ limit. In the next section, we will examine the leading corrections to the $d \to \infty$ limit and show that one still recovers an effective Markov process to leading order.

### 3. Large-d corrections

Away from the $d = \infty$ limit, the permutations are no longer completely locked to each other. The statistical mechanics model remains ferromagnetic and favours aligned permutations, but there are now corrections due to dilute formations of non-identity domains on top of the fully polarized state (the ferromagnetic ground state at $d = \infty$). A domain wall between permutations at different times (i.e., separated by a propagation step) incurs a free energy cost of $\log(d)$, whereas a domain wall within a collision step incurs a free energy cost of $12\log(d)$. At large $d$, the least costly domains are then given by the configurations shown in Fig. 2.

In analogy with typical first order perturbation theory, these configurations can be perturbatively accounted for by projecting the two copy Haar average collision gate onto the identity permutation subspace (i.e., onto the manifold of states accessed by the $d = \infty$ model). That is to say, we compute $G^{(2)} \equiv P_{\mathbb{1}}\mathbb{E}[U_C \otimes U_C \otimes U_C^* \otimes U_C^*]P_{\mathbb{1}}$, where the identity subspace projector $P_{\mathbb{1}}$ acts on the six links associated with each collision gate, $P_{\mathbb{1}} = \prod_{\ell=1}^{6} P_{\mathbb{1}}^{(\ell)}$. For a given link, the identity permutation projector is given by

$$P_{\mathbb{1}}^{(\ell)} = \sum_{n_1^{(\ell)}, n_2^{(\ell)}} \frac{1}{d^2} \left| n_1^{(\ell)}, n_2^{(\ell)}; \mathbb{1} \right\rangle \left\langle n_1^{(\ell)}, n_2^{(\ell)}; \mathbb{1} \right|. \tag{16}$$

The projector $P_{\mathbb{1}}$ has an obvious action on identity permutation states, i.e., $P_{\mathbb{1}}|a,b;\mathbb{1}\rangle = |a,b;\mathbb{1}\rangle$, but has a non-trivial action on the states $|a,b;\text{Sw}\rangle$ appearing in Eq. 7. Acting $P_{\mathbb{1}}$ on the state $|a,b;\text{Sw}\rangle$ only gives a non-zero answer when $a = b$. To see this, notice that in states in the identity permutation subspace, the two conjugate replicas must be in the same particle sector as their un-conjugated pair, i.e., $(a, a^*, b, b^*)$. Whereas for the swap states the identification is instead $(a, b^*, b, a^*)$. If $a \neq b$, it is not possible for these two vectors to be equal and therefore the overlap of these states must be zero. Therefore, we only need to compute the action of $P_{\mathbb{1}}$ on $|a,a;\text{Sw}\rangle$. Using the definition of this state in Eq. 6, and the definition of the identity projector, we find

$$P_{\mathbb{1}}|a,a;\text{Sw}\rangle = \sum_{\boldsymbol{n} \in a} \frac{1}{d^6} |\boldsymbol{n}, \boldsymbol{n}, \mathbb{1}\rangle. \tag{17}$$

For the non-colliding sectors $a \in \{\perp_k\}$, these subspaces are one-dimensional (excluding the ancilla degrees of freedom) so that the sum over particle configurations $\boldsymbol{n}$ is just a single term, this gives $P_{\mathbb{1}}|a,a;\text{Sw}\rangle = \frac{1}{d^6}|a,a,\mathbb{1}\rangle$. By Inserting the identity subspace projector either side of Eq. 7, collecting all of the $a,b \in \{\perp_k\}$ terms, and using this last result, we find that the action of $G^{(2)}$ on these non-colliding states is just given by the infinite $d$ action.

Lastly, we deal with case of $a = b \in 2, 3$. For $a = 2$ there are three particle configurations which we enumerate as $\boldsymbol{n} \in \{2_1, 2_2, 2_3\}$. Using this notation, we have $P_{\mathbb{1}}|2,2;\text{Sw}\rangle = \frac{1}{d^6}(|2_1, 2_1, \mathbb{1}\rangle) + |2_2, 2_2, \mathbb{1}\rangle) + |2_3, 2_3, \mathbb{1}\rangle)$. Similarly, for $a = 3$, we enumerate the two particle configurations as $\boldsymbol{n} \in \{3_1, 3_2\}$ and write $P_{\mathbb{1}}|3,3;\text{Sw}\rangle = \frac{1}{d^6}(|3_1, 3_1, \mathbb{1}\rangle) + |3_2, 3_2, \mathbb{1}\rangle)$. Since we have projected the two-copy averaged gate onto the identity permutation subspace, the permutation label is now redundant and will be dropped. Putting this all together gives the effective two-copy averaged gates as

$$G^{(2)} = \mathbb{E}[U_C \otimes U_C^*] \otimes \mathbb{E}[U_C \otimes U_C^*] + \frac{1}{d_2^2 - 1}|\psi_2\rangle\langle\psi_2| + \frac{1}{d_3^2 - 1}|\psi_3\rangle\langle\psi_3|, \tag{18}$$

where the states $|\psi_2\rangle$ and $|\psi_3\rangle$ are given by

$$|\psi_2\rangle = \frac{1}{d_2}|2_1 + 2_2 + 2_3, 2_1 + 2_2 + 2_3\rangle - \frac{1}{d^6}(|2_1, 2_1\rangle + |2_2, 2_2\rangle + |2_3, 2_3\rangle), \tag{19}$$

$$|\psi_3\rangle = \frac{1}{d_3}|3_1 + 3_2, 3_1 + 3_2\rangle - \frac{1}{d^6}(|3_1, 3_1\rangle + |3_2, 3_2\rangle). \tag{20}$$

Note that the first term corresponds to the decoupled single copy Haar averages while the remaining terms are the corrections coming from the dilute small domain wall configurations in Fig. 2.

### 4. Large-d effective Markov process

Eq. 18 defines the action of the two copy average within the identity permutation subspace. As we saw in the infinite $d$ limit, the ancilla degrees of freedom also factorize out here. By marginalizing over the ancilla degrees of freedom at the final time, we are contracting the transfer matrix by a factorized state $|\psi_{\text{particles}}\rangle \otimes |\mathbb{1}_{\text{ancilla}}\rangle$. The identity state for the ancilla can then simply be propagated through and contracted against the initial state, thus

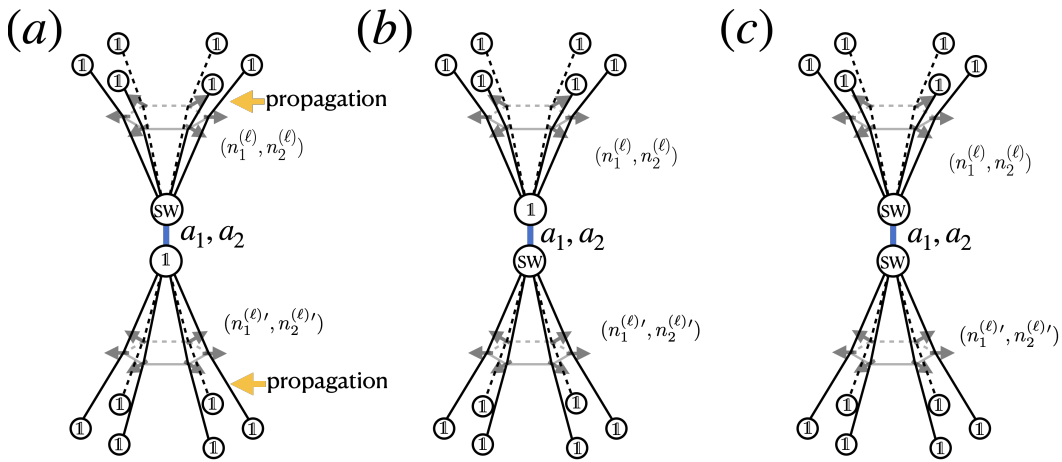

FIG. 2. **Minimal Cost Configurations.** (a) The swap permutation causes a difference between six inter-gate permutation states at time $t+1$ leading to a free energy cost of $6\log d$. The difference between the intra-gate permutation states leads to a free energy cost of $6\log d$. Thus the total cost for this domain is $12\log d$. (Here we have only given the leading order free energy costs at large $d$.) (b) Similar to (a) one has six inter-gate permutation differences with the permutations at $t-1$ and has one intra-gate permutation difference which again leads to a cost of $12\log d$. (c) In this scenario, corrections only come from inter-gate permutation differences. Since the vertices of internal (thick) edge carry the same permutations, the cost for this non-identity domain only comes from the disagreement between the permutations at $t+1$ and $t-1$, giving a cost of $12\log d$.

removing the ancilla from the transfer matrix calculation entirely. We are left with an effective evolution of the charge degrees of freedom within the identity permutation subspace.

For the resulting effective Markov process we will denote the classical states with curved brackets rather than the angled brackets used for the statistical mechanics model – $|\alpha,\beta)$ is the state in which the first (second) replica is in the particle configuration $\alpha$ ($\beta$). When the particle configurations in the replicas are in different collision subspaces the states evolve according to the decoupled single copy Markov process. For the case when the configurations are in the same collision subspace, the probabilities for transitions are modified.

Denote $\lambda_2 = \frac{1}{d_2^2-1}$ and $\lambda_3 = \frac{1}{d_3^2-1}$. Then the modified three body collision transitions are given by,

$$|\alpha,\alpha) \to \frac{1+\lambda_3}{4} \sum_{|\alpha')\in\mathcal{H}_3} |\alpha',\alpha') + \frac{1-\lambda_3}{4} \sum_{|\alpha'),|\beta')\in\mathcal{H}_3,\alpha'\neq\beta'} |\alpha',\beta'), \tag{21}$$

$$|\alpha,\beta) \to \frac{1-\lambda_3}{4} \sum_{|\alpha')\in\mathcal{H}_3} |\alpha',\alpha') + \frac{1+\lambda_3}{4} \sum_{|\alpha'),|\beta')\in\mathcal{H}_3,\alpha'\neq\beta'} |\alpha',\beta'), \tag{22}$$

where the latter equation only refers to the case $\alpha\neq\beta$. Finally, the modified two body collision transitions are given by,

$$|\alpha,\alpha) \to \frac{1+4\lambda_2}{9} \sum_{|\alpha')\in\mathcal{H}_2} |\alpha',\alpha') + \frac{1-2\lambda_2}{9} \sum_{|\alpha'),|\beta')\in\mathcal{H}_2,\alpha'\neq\beta'} |\alpha',\beta'), \tag{23}$$

$$|\alpha,\beta) \to \frac{1-2\lambda_2}{9} \sum_{|\alpha')\in\mathcal{H}_2} |\alpha',\alpha') + \frac{1+\lambda_2}{9} \sum_{|\alpha'),|\beta')\in\mathcal{H}_2,\alpha'\neq\beta'} |\alpha',\beta'), \tag{24}$$

where once again the latter transition only refers to the case $\alpha\neq\beta$. Here we have slightly abused notation for $\mathcal{H}_a$ to only refer to the subspace of two and three-body collision configurations. By inspection one can see that the above coefficients sum to one and that they are positive. Thus we found that an effective Markov process governs the two copy dynamics in the large-$d$ limit.

## II. DERIVATION OF LATTICE GAS AUTOMATA HYDRODYNAMICS

In this section we derive the hydrodynamics of the random quantum circuit. Since we will be dealing with the behavior of expectation values of operators the derivation closely follows the derivation in Ref [6]. In this section we

will adopt the shorthand, $n_i(\vec{x}, t) \equiv \langle n_{\vec{x}}^i(t+1) \rangle$. For a single circuit realization, we have

$$n_i(\vec{x} + \hat{b}_i, t+1) = n_i(\vec{x}, t) + \Omega_i(\{n_j(\vec{x}, t)\}_{j=1}^6), \tag{25}$$

where $\Omega_i(\{n_j(\vec{x}, t)\}_{j=1}^6)$ is a collision factor which encapsulates the result of the collision update based on the neighboring links of $n_i(\vec{x}, t)$.

We will expand the left hand side to second order in space and first order in time to capture up to diffusive corrections for the hydrodynamics. One arrives at

$$\partial_t n_i + b_i^j \partial_j n_i + \frac{1}{2} b_i^j b_i^k \partial_j \partial_k n_i = \Omega_i(\{n_j(\vec{x}, t)\}_{j=1}^6), \tag{26}$$

where employed Einstein summation for the indices not equal to $i$. As done in the main text we defined coarse grained density and "momentum" variables, given by $n(\vec{x}, t) = \sum_i n_i(\vec{x}, t)$ and $n(\vec{x}, t) v(\vec{x}, t) = \sum_i \hat{b}_i n_i(\vec{x}, t)$. At the level of operators one has that,

$$\sum_i n_{\vec{x}}^i(t) = \sum_i n_{\vec{x}_i + \hat{b}_i}^i(t+1), \tag{27}$$

$$\sum_i \hat{b}_i n_{\vec{x}}^i(t) = \sum_i \hat{b}_i n_{\vec{x}_i + \hat{b}_i}^i(t+1). \tag{28}$$

This can be seen by noting that the total particle number for a given hexagonal cell $\sum_i n_{\vec{x}}^i(t)$ and $\sum_i \hat{b}_i n_{\vec{x}}^i(t)$ are invariant under the collision gates. One can show that the propagation gates take $n_{\vec{x}}^i(t) \to n_{\vec{x}+\hat{b}_i}^i(t)$ since the $n_{\vec{x}}^i$ are diagonal in the occupation basis and the propagation gates are comprised of SWAP gates which send product states in the occupation basis to product states.

By taking the expectation value of both sides of Eq. 27 and Eq. 28, one can see that $\sum_i \Omega_i(\{n_j(\vec{x}, t)\}_{j=1}^6) = 0$ and $\sum_i \hat{b}_i \Omega_i(\{n_j(\vec{x}, t)\}_{j=1}^6) = 0$. Using this observation we arrive at

$$\partial_t n + b_i^j \partial_j n_i + \frac{1}{2} b_i^j b_i^k \partial_j \partial_k n_i = 0, \tag{29}$$

$$\partial_t b_i^l n_i + b_i^j \partial_j b_i^l n_i + \frac{1}{2} b_i^j b_i^k \partial_j \partial_k b_i^l n_i = 0, \tag{30}$$

where we are now employing Einstein summation convention over all indices.

We now expand $n_i$ about its local equilbrium value characterized by conjugate thermodynamic parameters $\vec{u}(\vec{x}, t)$ associated with "momentum" conservation and $\mu(\vec{x}, t)$, the local chemical potential. Furthermore, we assume that the mean velocity $v_i(\vec{x}, t)$ is small compared to the speed of sound of the system, $c$.

To leading order

$$n_i(\vec{x}, t) = n_i^0(\vec{x}, t) = \frac{1}{1 + e^{\mu + \vec{u} \cdot \hat{b}_i}}. \tag{31}$$

We can expand the above equations up to second order in $|\vec{v}|$ or equivalently to second order in $|\vec{u}|$. One finds that the local equilibrium value up to second order in $|\vec{v}|$ is given by,

$$n_i(\vec{x}, t) \simeq \frac{n}{6} + \frac{n}{3} b_i^k v_k + \frac{n}{3} \frac{6-2n}{6-n} \left( b_i^j b_i^k - \frac{1}{2} \delta^{jk} \right) v_j v_k. \tag{32}$$

We can add corrections, denoted by $n_i^1(\vec{x}, t)$ beyond the local equilibrium value, which correspond to adding gradients of the density and "momentum". The general lowest order form such a term can take is,

$$n_i^1(\vec{x}, t) = \alpha b_i^j \partial_j n + (\beta b_i^j b_i^k + \gamma \delta^{jk}) \partial_j n v_k. \tag{33}$$

To leading order $\sum_i n_i^1 = 0$ and $\sum_i \hat{b}_i n_i^1 = 0$. Using $\sum_i b_i^j = 0$ and $\sum_i b_i^j b_i^k = 3\delta^{jk}$, one finds that $\alpha = 0$ and $\beta + 2\gamma = 0$.

Substituting the full expression for $n_i(\vec{x}, t) = n_i^0(\vec{x}, t) + n_i^1(\vec{x}, t)$ into Eq. 29 and Eq. 30, one finds that

$$\partial_t n + \partial_j n v_j = 0, \tag{34}$$

$$\partial_t n v_i + A_{ijkl} \partial_j (c^2 n g v_k v_l) = -\partial_i P + (\eta - \frac{1}{8}) A_{ijkl} \partial_j \partial_k n v_l, \tag{35}$$

7

where $c^2 = 1/2$, $g(n) = \frac{3-n}{6-n}$, $P = c^2 n$ and $A_{ijkl} = \delta_{il}\delta_{jk} + \delta_{ik}\delta_{jl} - \delta_{ij}\delta_{kl}$. We note that $\eta = -x\frac{3}{4}\beta$.

We can identify the momentum stress tensor and write the above equations as,

$$\partial_t n + \partial_j n v_k = 0, \tag{36}$$
$$\partial_t n v_i + \partial_j \langle \pi_{ij} \rangle = 0, \tag{37}$$

with $\langle \pi_{ij} \rangle = P\delta_{ij} + A_{ijkl} c^2 \rho g v_k v_l + (\eta - \frac{1}{8}) A_{ijkl} \partial_k n v_l$.

## III. NUMERICAL SIMULATIONS OF AC SHEAR VISCOSITY

Here we outline how the numerics were performed for computing the AC Shear viscosity, $\eta(\omega)$. Recall $\eta(\omega)$ is given by the Kubo formula,

$$\eta(\omega) = \frac{1}{2}C(0) + \sum_{t=1}^{\infty} e^{i\omega t} C(t), \tag{38}$$

where $C(t) = \frac{1}{V}\langle \Pi_{xy}(t) \Pi_{xy}(0) \rangle$ with $V$ the volume of the system, $\Pi_{ij} = \sum_{\vec{x}} \pi_{ij}(\vec{x})$, $\langle A \rangle = \text{tr}(A\rho_{\text{eq}})$ and $\rho_{\text{eq}} \propto e^{-\mu N}$ where $N$ denotes the total particle number operator.

To compute the variance we must evaluate both $\mathbb{E}[\eta(\omega)]$ and $\mathbb{E}[|\eta(\omega)|^2]$. The former can be simulated classically since the single copy average corresponds to the variant of the FHP model. More explicitly we can evaluate $\mathbb{E}[C(t)]$ by sampling (classical) configurations on the triangular lattice according to $\rho_{\text{eq}}$.

One updates these states by first applying the collision single copy transfer matrix gates which corresponds to changing configurations in the two body and three body collision subspaces with probabilities $1/2$ and $1/3$, respectively. Then one propagates degrees of freedom one lattice spacing in the direction their arrow points as outlined in the main text. This process achieves one full time step. One then averages over all initial states and evolutions to obtain $\mathbb{E}[C(t)]$.

We also need to evaluate $\mathbb{E}[|\eta(\omega)|^2]$. Expanding this term, one has

$$\mathbb{E}[|\eta(\omega)|^2] = \frac{1}{4}C(0)^2 + \mathbb{E}[\text{Re}(\sum_{t=1}^{\infty} e^{i\omega t} C(t) C(0))] + \mathbb{E}\left[\left|\sum_{t=1}^{\infty} e^{i\omega t} C(t)\right|^2\right]. \tag{39}$$

The first term can be evaluated by sampling from $\rho_{\text{eq}}$ and the second term is evaluated by simulating the single copy Markov process as previously discussed. The last term requires simulating both the single copy and two copy dynamics.

This can be seen by expanding the last term out, one has

$$\mathbb{E}\left[\left|\sum_{t=1}^{\infty} e^{i\omega t} C(t)\right|^2\right] = \sum_{t,t'=1}^{\infty} e^{i\omega(t-t')} \mathbb{E}[C(t)C(t')]. \tag{40}$$

In the case when $t = t'$ evolves only with the two copy averaged dynamics as the number of unitary operators which appear in $C(t)$ and $C(t')$ is the same. However when $t < t'$, then only has paired unitary operators between the replicas up until time $t$. This means that one first evolves with the two copy averaged dynamics until time $t$ and then one switches over to evolving configurations on the (without loss of generality) second copy until time $t'$.

In the simulations we carried out, we averaged over $8.9 \times 10^6$ realizations and used a chemical potential value of $\mu = 0.5$.

---


\end{document}